\documentclass[10pt,oneside]{amsart}

\usepackage{amsmath}    
\usepackage{graphicx}   
\usepackage{amssymb}
\usepackage[latin1]{inputenc}     
\usepackage{float}  

      \theoremstyle{plain}
      \newtheorem{theorem}{Theorem}[section]
      \newtheorem{lemma}[theorem]{Lemma}

      \theoremstyle{definition}
      \newtheorem{definition}[theorem]{Definition}

      \theoremstyle{remark}
      \newtheorem{remark}[theorem]{Remark}

      \makeatletter
      \def\@setcopyright{}
      \def\serieslogo@{}
      \makeatother

\begin{document}



   \author{Michel Gondran}
   \address{EDF R\S D, 1 av.du Général de Gaulle, 92140 Clamart,France}
   \email{michel.gondran@edf.fr}

\title{Schrödinger Proof in Minplus complex Analysis}

\begin{abstract}
We are presenting an internal trajectory model for a quantum
particle in the Schrödinger non-relativistic approximation. This
model is based on two new mathematical concepts: a complex
analytical mechanics in Minplus complex analysis and a periodical
non random process which gives a complex Îto formula.

This model naturally generates a concept of spin or isospin and
the Heisenberg inequalities, and leads to the Schrödinger equation
using a generalization of the least action principle adapted to
the trajectories of this type.
\end{abstract}

   \subjclass{Primary 49505; Secondary 81Q05, 81V22}

   \keywords{Schrödinger equation, complex, idempotent analysis.}

   \thanks{This work was partially supported by the Erwin Schrödinger
    International Institute for Mathematical Physics (ESI)}


   \date{\today}


   \maketitle

\section{Introduction}

\noindent

\bigskip We are presenting a complex process to model a quantum particle in
the Schrödinger non-relativistic approximation. This model
naturally generates an intrinsic angular momentum and leads to the
Schrödinger equation using a generalization of the least action
principle.

The model is based on two new mathematical concepts that we have
brought up in previous papers: a complex analytical mechanics
\cite{Gondran2},\cite{GondranHoblos} and a periodical
deterministic process \cite{Gondran3}.

The purpose of this paper is to develop this new model and to
demonstrate some properties. We attempt to address the following
questions: how does an intrinsic angular momentum emerge from the
six periodical complex processes (theorem 2.7)? How can we find
the Heisenberg inequalities (theorem 2.8)? the non commutation
relations (theorem 2.9)? On what principle can the Schrödinger
equation be demonstrated (theorems 3.4 and 3.5)? How can we find
the trajectories of de Broglie and Bohm (theorem 3.6)?

\section{The model of trajectory}

\noindent
In any orthonomal reference system R, let be U$_{R}$ the set of 8 vectors $%
\left\{ e_{1},e_{2},e_{3}\right\} $ where e$_{i}=\pm 1$, corresponding to
the eight vertices of a cube. We then consider the set S of the circular
permutations of the 6 vertices of this cube to correspond to a sequence of
adjacent vertices. We verify that S corresponds to the following eight
permutations:
\begin{figure}[H]
\begin{center}
\includegraphics[width=0.9\linewidth]{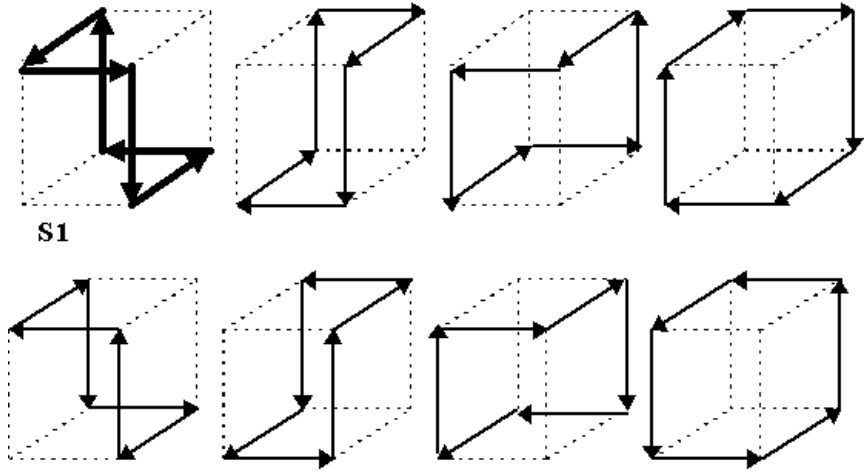}

\caption{\label{fig:spin}The eight permutations.}
\end{center}
\end{figure}
To each of these permutations $s\in S$, we associate the set $%
U_{R}^{s}$ of these vertices. This gives $s^{6}u^{j}=u^{j}$ for any $%
u^{j}\in U_{R}^{s}$ .

\bigskip
\begin{remark}
Each of these permutations corresponds to the rotation of the cube
on a diagonal. Therefore, it will be necessary to choose one of
these permutations in the case of the Schrödinger equation. In the
case of the Pauli equation, the frame R will no longer be fixed
but will be oriented along an axis (that of the spin or isospin)
which will change with time under the effect of a magnetic field.
\end{remark}

\bigskip

\begin{definition}
\textit{For any time step $\varepsilon
>0$ and for any permutation $s\in S$, we define the 6 discrete
processes $Z_{\varepsilon
}^{j}(t)$ at time $t=n\varepsilon ,$ with $n=6q+r$ ($n,q,r$ integers and $%
0\leq r\leq 5$), by: }
\begin{equation}
Z_{\epsilon }^{j}(\left( n+1\right) \epsilon )=Z_{\epsilon }^{j}(n\epsilon )+%
\mathcal{V}(6(q+1)\epsilon )\epsilon +\gamma
(s^{n+1}u^{j}-s^{n}u^{j})
\end{equation}

\begin{equation}
Z_{\epsilon }^{j}(0)=Z_{0},\qquad u^{j}\in U_{R}^{s}
\end{equation}
\textit{where $\gamma =$ $(1+i)\sqrt{\frac{3\hslash \varepsilon}{8m}}$ and where $%
\mathcal{V}(t)$ corresponds to a continuously differentiable
complex function, where $\hslash $ is the Planck constant and m is
the mass of the particle, and where $Z_{0}$ is a given vector of
$\mathcal{C}^3$.}

\end{definition}

\bigskip
\begin{remark}
The 6 processes $Z_{\varepsilon }^{j}(t)$ look like stochastic
processes of Nelson \cite{Nelson1},\cite{Nelson2}, but there are
very different because there are deterministic. In $\left[
6\right] $ p.117, Feynman and Hibbs show that the ''important
paths'' of quantum mechanics, although continuous, are very
irregular and nowhere differentiable. They admit an average
velocity (lim$_{\Delta t\rightarrow 0^{+}}\left\langle
\frac{x_{k+1}-x_{k}}{\Delta t}\right\rangle =v)$, but no average
quadratic velocity
$\left( \left\langle \left( \frac{%
x_{k+1}-x_{k}}{\Delta t}\right) ^{2}\right\rangle =\frac{i\hbar }{m\Delta t}%
\right) .$ The processes $Z_{\varepsilon }^{j}(t)$ are built on
the same idea, but with an $\varepsilon =\Delta t$ very small but
\textit{finite} and with an \textit{strong relation between the 6
processes}.
\end{remark}

\bigskip

We denote $\widetilde{Z}_{\varepsilon }\left( t\right) $ the solution of the
discrete system defined at time $t=n\varepsilon $ with $n=6q+r$ ($n,q,r$
integers and $0\leq r\leq 5$), by:

\begin{equation}
\widetilde{Z}_{\epsilon }(\left( n+1\right) \epsilon )=\widetilde{Z}%
_{\epsilon }(n\epsilon )+\mathcal{V}(6(q+1)\epsilon )\epsilon
\end{equation}

\begin{equation}
\widetilde{Z}_{\epsilon }(0)=Z_{0}.
\end{equation}

At all times $n$ $\epsilon ,$ we verify that:

\begin{equation}
Z_{\epsilon }^{j}(n\epsilon )=\widetilde{Z}_{\varepsilon }(n\epsilon )+(1+i)%
\sqrt{\frac{3\hslash \varepsilon }{8m}}\left( s^{n}u^{j}-u^{j}\right) .
\end{equation}

As $s^{6}u^{j}=u^{j}$, we deduce from $\left( 5\right) $ that for all $j$, $%
Z_{\varepsilon }^{j}(6q\varepsilon )=\widetilde{Z}_{\varepsilon }\left(
6q\varepsilon \right) .$ \bigskip Then, we have for all $j$ and $t$, $%
Z_{\epsilon }^{j}(t)=\widetilde{Z}_{\varepsilon }\left( t\right) +0\left(
\sqrt{\varepsilon }\right) .$

We denote $\widetilde{Z}\left( t\right) $, the solution of the classical
differential equation

\begin{equation}
\frac{d\widetilde{Z}(t)}{dt}=\mathcal{V}(t)
\end{equation}

\begin{equation}
\text{\ \ }\widetilde{Z}(0)=Z_{0}.
\end{equation}

Because $\mathcal{V}(t)$ is continuously differentiable, we have for each $%
t=n\varepsilon $, $\widetilde{Z}_{\varepsilon }\left( t\right) =$ $%
\widetilde{Z}\left( t\right) +0\left( \varepsilon \right) $ and thus $%
Z_{\epsilon }^{j}(t)=\widetilde{Z}\left( t\right) +0\left( \sqrt{\varepsilon
}\right) .$We conclude , when $\varepsilon \rightarrow 0^{+},$ that each $%
Z_{\epsilon }^{j}(t)$ converges to $\widetilde{Z}\left( t\right) .$

Equation $\left( 5\right) $ shows that the 6 processes
$Z_{\epsilon }^{j}(t)$ correspond to processes which oscillate
around $\widetilde{Z}_{\varepsilon }\left( t\right) $ and which
are, as the "Feynman paths", more and more irregular and nowhere
differentiable as $\varepsilon \rightarrow 0^{+}.$ We call
$\widetilde{Z}_{\varepsilon }\left( t\right) $
\textit{the basic trajectory}. This is the mean of the 6 processes $%
Z_{\epsilon }^{j}(t).$

The 6 positions of processes $Z_{\varepsilon }^{j}(t)$ at times $\
t=6q\varepsilon $ correspond to only one position on the basic trajectory $%
\left( Z_{\varepsilon }^{j}(6q\varepsilon )=\widetilde{Z_{\varepsilon }}%
\left( 6q\varepsilon \right) \right) $. The process defined by $\left(
1\right) $ and $\left( 2\right) $ leaves the basic trajectory of $%
6q\varepsilon $ to $\left( 6q+1\right) \varepsilon $ and returns to the
basic trajectory between $\left( 6q+5\right) \varepsilon $ and $\left(
6q+6\right) \varepsilon $.

\bigskip
\begin{definition}
 \textit{We denote
$Z_{\epsilon }(t)$ the discrete process forming each of the six
complex processes $Z_{\epsilon }^{j}(t)$ with a weight of
$\frac{1}{6}$. We associate this process to a particle. Then, a
particle is represented by six processes. Then
$\widetilde{Z}_{\varepsilon }\left( t\right) $ can be interpreted
as to the complex mean position of the particle.}

\end{definition}

\bigskip

A possible interpretation of the imaginary part of processes $Z_{\varepsilon
}^{j}(t)$ corresponds to a bivector of the Clifford algebra $Cl_{3}.$ We
discuss this point in conclusion.

We note $X_{\varepsilon }^{j}\left( t\right) $ the real part of process $%
Z_{\varepsilon }^{j}(t)$. Equation (5) therefore becomes:

\begin{equation}
{\large X}_{\varepsilon }^{j}(n\varepsilon )=\widetilde{{\large X}}%
_{\varepsilon }(n\varepsilon )+\sqrt{\frac{3\hslash \varepsilon }{8m}}%
(s^{n}u^{j}-u^{j})
\end{equation}
\begin{figure}[H]
\begin{center}
\includegraphics[width=0.9\linewidth]{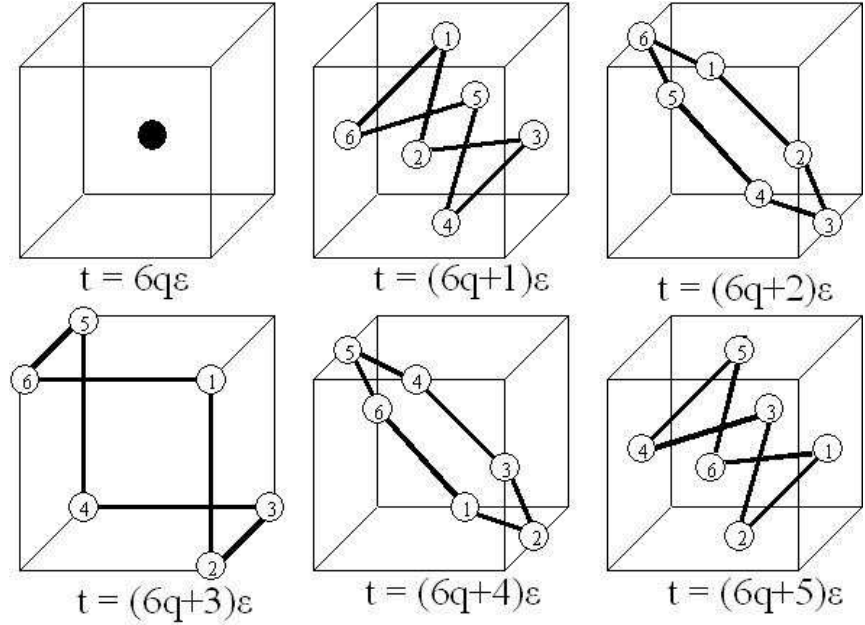}

\caption{\label{fig:evolution}Evolution of the six processes
$X_\varepsilon^j(n\varepsilon)$.}
\end{center}
\end{figure}
The evolution of the six processes ${\large X}_{\varepsilon
}^{j}(n\varepsilon )$ visualizes in the figure 2. At time
$t=6q\varepsilon$, the six points are in the center of a cube. At
times $t=(6q+1)\varepsilon$ and $t=(6q+5)\varepsilon$, the six
points are in the center of the six faces of this cube. At times
$t=(6q+2)\varepsilon$ and $t=(6q+4)\varepsilon$, the six points
are on a circle in the middle of six edges of the cube. At time
$t=(6q+3)\varepsilon$, the six points are on the six vertices of
the cube.

\bigskip

\begin{remark}
It is possible to give more interpretations of the processes $%
Z_{\varepsilon }^{j}(t)$: non standard processes, fiber space,
string. In the first interpretation, each process $%
Z_{\varepsilon }^{j}(t)$ is a non standard process in spite of the
non standard analysis of Robinson \cite{Robinson} (field
$^\varepsilon \mathbb{R}$ of Robinson built with the infinitesimal
$\varepsilon$ added to $\mathbb{R}$) with $\widetilde{Z}\left(
t\right)$ as standard part. In the fiber space
interpretation, the process $%
Z_{\varepsilon }^{j}(t)$ corresponds to \textit{the total space}
and $\widetilde{Z}_{\varepsilon }\left( t\right) $ corresponds to
\textit{the basic space}.

The third interpretation of process defined by $%
\left( 1\right) $ and $\left( 2\right) $ is to consider this
process as a \textit{special elastic string model}. Its length at
times $t=6q\varepsilon $ is equal to zero. At times $t\neq
6q\varepsilon $, it takes a finite length and the 6 points
$X_{\varepsilon }^{j}(t)$ correspond to 6 points of a string. The
motion of these points therefore corresponds to a vibration of the
string. Moreover this interpretation suggests a sort of creation
process between times $t=6q\varepsilon $ and $\left( 6q+1\right)
\varepsilon $ followed by
an annihilation process between $\left( 6q+5\right) \varepsilon $ and $%
6\left( q+1\right) \varepsilon .$

\end{remark}

\bigskip

\begin{definition}
\textit{The mean angular momentum of a process following (8) is
given by:}

\[
\sigma =E_{n,j}\left( \sigma _{n}^{j}\right) =\frac{1}{36}%
\sum_{n=6q}^{6q+5}\sum_{j=1}^{6}\sigma _{n}^{j}\text{ \ \
\textit{with}}
\]
\[
\sigma _{n}^{j}=r_{n}^{j}\wedge p_{n}^{j}\text{ , \ }r_{n}^{j}={\large X}%
_{\varepsilon }^{j}(n\varepsilon
)=\widetilde{r}_{n}+\sqrt{\frac{3\hslash \varepsilon }{8m}}\left(
s^{n}u^{j}-u^{j}\right) \text{ }
\]
\[
\text{\textit{and }\ }p_{n}^{j}=m\frac{r_{n+1}^{j}-r_{n}^{j}}{\varepsilon }=m\frac{%
\widetilde{r}_{n+1}-\widetilde{r}_{n}}{\varepsilon }+\sqrt{\frac{3\hslash m}{%
8\varepsilon }}\left( s^{n+1}u^{j}-s^{n}u^{j}\right) .
\]

\end{definition}

 Using \ $\sum_{j=1}^{6}s^{n}u^{j}=0$ \ for all n, it follows
that
\[
\ \sigma =\widetilde{r}\wedge m\widetilde{v}+\frac{1}{16}\hslash
\sum_{j=1}^{6}\left( u^{j}\wedge su^{j}\right)
\]
with $\widetilde{r}=\frac{1}{6}\sum_{n=6q}^{6q+5}\widetilde{r}_{n}$ and $%
\widetilde{v}=\widetilde{v}\left( 6\left( q+1\right) \varepsilon \right) .$

For instance, for $s=s_{1}$, that is for the permutation defined by the
sequence of the 6 vertices: $u^{1}=\left\{ -1,-1,-1\right\} $, $%
u^{2}=\left\{ -1,-1,+1\right\} $, $u^{3}=\left\{ 1,-1,1\right\} $, $%
u^{4}=\left\{ 1,1,1\right\} $, $u^{5}=\left\{ 1,1,-1\right\} $, $%
u^{6}=\left\{ -1,1,-1\right\} $, we have
\[
\sum_{j=1}^{6}\left( u^{j}\wedge s_{1}u^{j}\right) =\left\{ -2,2,0\right\}
+\left\{ 0,2,2\right\} +\left\{ -2,0,2\right\} +\left\{ -2,2,0\right\}
\]
\[
+\left\{ 0,2,2\right\} +\left\{ -2,0,2\right\} =8\left\{ -1,1,1\right\}
\]

and then:

\[
\sigma _{x}=m\left( y\widetilde{v}_{z}-z\widetilde{v}_{y}\right) -\frac{%
\hslash }{2}\text{, \ }\sigma _{y}=m\left( z\widetilde{v}_{x}-x\widetilde{v}%
_{z}\right) +\frac{\hslash }{2}\text{ }
\]
\[
\text{ and \ \ }\sigma _{z}=m\left( x\widetilde{v}_{y}-y\widetilde{v}%
_{x}\right) +\frac{\hslash }{2}.
\]

We deduce the following theorem:

\bigskip

\begin{theorem}
\textit{\ For any }$\varepsilon >0$\textit{\ and for }$%
s=s_{1}$\textit{, the real part of process defined by }$\left(
1\right) \left( 2\right) $\textit{\ has a mean intrinsic angular
momentum of
components: }$s_{x}=+\frac{\hslash }{2},s_{y}=-\frac{\hslash }{2},s_{z}=-%
\frac{\hslash }{2}$\textit{. The 7 other combinations }$\left( \pm \frac{%
\hslash }{2}\right) $\textit{\ are given by the 7 other permutations of }$S$%
\textit{.}
\end{theorem}

\bigskip

Let be $\widetilde{x}_{\varepsilon }(n\varepsilon )$ the real mean position
on the $x$ axis of the particle at time $n\varepsilon $ (with $n=6q+r$) and $%
m\widetilde{v}=m\widetilde{v}\left( 6(q+1)\varepsilon \right) $
the real mean momentum. The calculation of standard deviations
$\triangle x$\ and $\Delta
p_{x}$ of the position and of the momentum on the x axis, at time $%
n\varepsilon $ with $n=6q+w$ $\ \left( \left| w\right| \leq 3\right) $,
yields:

\[
\left( \Delta x\right) _{n\varepsilon }=\sqrt{\frac{1}{6}\sum_{j}\left(
r_{n}^{j}-\widetilde{r_{n}}\right) _{x}^{2}}=\sqrt{\frac{\hslash \varepsilon
}{2m}}\sqrt{\left| w\right| }
\]

\[
\left( \Delta p_{x}\right) _{n\varepsilon }=\sqrt{\frac{1}{6}\sum_{j}\left(
p_{n}^{j}-\widetilde{p_{n}}\right) _{x}^{2}}=\sqrt{\frac{\hslash m}{%
2\varepsilon }}
\]

For exemple, we have for $s=s_{1}$ and $n=6q+1:$%
\[
\left( \Delta x\right) _{n\varepsilon }=\sqrt{\frac{3\hslash \varepsilon }{8m%
}}\sqrt{\frac{1}{6}\sum_{j}\left( s_{1}u^{j}-u^{j}\right) _{x}^{2}}
\]
\[
=\sqrt{\frac{3\hslash \varepsilon }{8m}}\sqrt{\frac{1}{6}\left(
2^{2}+2^{2}\right) }=\sqrt{\frac{\hslash \varepsilon }{2m}}
\]

and
\[
\left( \Delta p_{x}\right) _{n\varepsilon }=\sqrt{\frac{3\hslash m}{%
8\varepsilon }}\sqrt{\frac{1}{6}\sum_{j}\left(
s_{1}^{2}u^{j}-s_{1}u^{j}\right) _{x}^{2}}
\]
\[
=\sqrt{\frac{3\hslash m}{8\varepsilon }}\sqrt{\frac{1}{6}\left(
2^{2}+2^{2}\right) }=\sqrt{\frac{\hslash m}{2\varepsilon }}.
\]

We deduce the second theorem:

\bigskip

\begin{theorem}
\textit{For any }$\varepsilon >0$%
\textit{\ and for any }$s$\textit{, the real part of process defined by }$%
\left( 1\right) \left( 2\right) $\textit{\ verifies\ the
Heisenberg
inequalities at any point not located on the basic trajectory (i.e. for any }%
$\ t=n\varepsilon \neq 6q\varepsilon $\textit{):}

\begin{equation}
\left( \Delta x\right) _{n\varepsilon }\cdot \left( \Delta
p_{x}\right) _{n\varepsilon }\geq \frac{\hslash }{2}.
\label{Math21}
\end{equation}

\end{theorem}

\bigskip
Let us show that the non commutation relations of the type
\[
\widehat{p}_{x}x-x\widehat{p}_{x}=-i\hslash
\]
can be interpreted as the non temporal commutation of $p_{x}\left(
t\right) $ and $x\left( t\right) $. Accordingly, we calculate the
mean of $p_{x}\left(
t+\varepsilon \right) x\left( t\right) -x\left( t+\varepsilon \right) $ $%
p_{x}\left( t\right) $:
\[
E\left( p_{x}\left( t+\varepsilon \right) x\left( t\right)
-x\left( t+\varepsilon \right) p_{x}\left( t\right) \right) =
\]

\[
=E_{n,j}\left( 1+i\right) ^{2}\left( \left( p_{n+1}^{j}\right)
_{x}\left( r_{n}^{j}\right) _{x}-\left( r_{n+1}^{j}\right)
_{x}\left( p_{n}^{j}\right) _{x}\right)
\]
where $r_{n}^{j}$ and $p_{n}^{j}$ are defined in definition 3. We
then obtain the desired result:

\bigskip

\begin{theorem}
\textit{For any }$\varepsilon >0$\textit{%
\ and for any }$s$\textit{, the process }defined by $\left(
1\right) \left(
2\right) $\textit{\ verifies} \textit{the non temporal commutativity relation%
}

\[
E\left( p_{x}\left( t+\varepsilon \right) x\left( t\right)
-x\left( t+\varepsilon \right) p_{x}\left( t\right) \right)
=-i\hslash .
\]
\end{theorem}

\bigskip

Moreover, we verify also that

\[
E\left( p_{x}\left( t+\varepsilon \right) y\left( t\right)
-y\left( t+\varepsilon \right) p_{x}\left( t\right) \right) =0.
\]

\bigskip

Let be $f\left( x,t\right) $ an application of class $C^{2}$ of $\Bbb{C}%
^{3}\times \Bbb{R}$ in $\Bbb{C}$ and $Z_{\epsilon }^{j}\left(
t\right) $ a process defined by (1) and (2). We denote\textit{\
the complex Dynkin operator, }previously introduced by Nottale
\cite{Nottale} under the name of ''quantum covariant derivative'':
\begin{equation}
D=\frac{\partial }{\partial t}+\mathcal{V}\cdot \triangledown -i\frac{%
\hslash }{2m}\triangle
\end{equation}

\bigskip

\begin{lemma}
\textsc{\ }\textit{For any }$\varepsilon
>0$\textit{\
and for any }$s$, the\textit{\ process }$Y_{\varepsilon }\left( t\right) $%
\textit{\ defined by}

\begin{equation}
Y_{\epsilon }\left( t\right) =E_{j}f(Z_{\epsilon }^{j}\left( t\right) ,t)=%
\frac{1}{6}\ \sum_{j}\left( f\left( Z_{\epsilon }^{j}\left(
t\right) ,t\right) \right)
\end{equation}
\textit{with }$Z_{\epsilon }^{j}\left( t\right) $\textit{\ based
on (1) and (2), verifies for any integer }$q$\textit{:}

\begin{equation}
Y_{\epsilon }\left( 6q\epsilon \right) -Y_{\epsilon }\left( \left(
6q-1\right) \epsilon \right) =Df\left( \widetilde{Z}\left(
6q\epsilon \right) ,6q\epsilon \right) \epsilon +o \left( \epsilon
\right).
\end{equation}

\end{lemma}

\textit{Proof}: First, we have $Y_{\epsilon }\left( 6q\epsilon \right)
=f\left( \widetilde{Z}\left( 6q\epsilon \right) ,6q\epsilon \right) $. Using
$\left( 5\right) $ and $\sum_{j}s^{n}u^{j}=0$, we find for all $t=n$ $%
\epsilon $

\[
E_{j}f(Z_{\epsilon }^{j}\left( t\right) ,t)=f(\widetilde{Z}_{\epsilon
}(t),t)+
\]

\[
+\frac{3i\hslash }{8m}\epsilon E_{j}\left\{ \sum_{k,l}\frac{\partial ^{2}f}{%
\partial x_{k}\partial x_{l}}\left( s^{n}u^{j}-u^{j}\right) _{k}\left(
s^{n}u^{j}-u^{j}\right) _{l}\right\} +o \left( \epsilon \right) .
\]

For $n=6q-1$,
\[
E_{j}\left( s^{n}u^{j}-u^{j}\right) _{k}\left( s^{n}u^{j}-u^{j}\right) _{l}=%
\frac{4+4}{6}\delta _{kl}
\]
and the calculation of the last term of $E_{j}f(Z_{\epsilon }^{j}\left(
t\right) ,t)$ gives $\frac{3i\hslash }{8m}\epsilon \frac{8}{6}\Delta f$.
Then we deduce

\[
Y_{\epsilon }\left( (6q-1)\epsilon \right) =f(\widetilde{Z}_{\epsilon
}((6q-1)\epsilon ),(6q-1)\epsilon )+
\]

\[
+i\frac{\hslash }{2m}\epsilon \Delta f(\widetilde{Z}_{\epsilon
}((6q-1)\epsilon ),(6q-1)\epsilon )+o \left( \epsilon \right) .
\]

Hence, with the development to first order of $f(\widetilde{Z}_{\epsilon
}((6q-1)\epsilon ),(6q-1)\epsilon )$, equation $\left( 12\right) $ follows. $%
\Box $

\bigskip
\begin{remark}
The process defined by $\left( 1\right) $ and $\left( 2\right) $
is only an example of processes that we could build with
permutation groups about $U_{R}$. We can preserve the
previous properties, if we change the permutation $s$ or the step of time $%
\varepsilon $, between 2 positions of the basic trajectory. We
will use this change in $\varepsilon $ in the remark 3.7 where we
give a possible interpretation of $\varepsilon .$

The second part of the process defined by $\left( 1\right) $ is
discrete. There are many solutions to make this part continuous
and differentiable, for example by a trajectory differentiable on
the circumscribed sphere which passes through all the 6 vertices
of each discrete process.
\end{remark}

\bigskip

\section{The Schrödinger equation}

\bigskip

We then construct a \textit{complex analytical mechanics} in the
same way as the conventional analytical mechanics but with objects
having a complex position $Z(t)\in \Bbb{C}^{3}$, a complex
velocity $\mathcal{V}\left( t\right) \in \Bbb{C}^{3}$ and using
the minimum of a complex function and complex minplus analysis, as
introduced in \cite{Gondran2}, \cite{GondranHoblos}. It is a
generalisation for the complex functions of the idempotent
analysis introduced by Maslov \cite{Maslov}. We recall the basic
ideas in the two following definitions.

\bigskip
\begin{definition}
\textit{Let a complex function $f\left( z\right) =f\left(
x+iy\right) $ from $\ \Bbb{C}^{n}$ to $\Bbb{C}$ such as $f\left(
z\right) =P\left( x,y\right) +iQ\left( x,y\right) $ with $P\left(
x,y\right) $
continuous at $x$ and $y$, and a closed set $A\subset \Bbb{C}^{n}$ such as $%
A=\left\{ x+iy/x\in X\subset \Bbb{R}^{n},y\in Y\subset
\Bbb{R}^{n}\right\} .$ We define \textit{the minimum of }$f$ for
$z\in A$, if it exists, by:
\[
\min_{z\in \mathit{A}}{f\left( z\right)}=\left\{ f\left(
z_{0}\right) \right\}
\]
where $z_{0}=x_{0}+iy_{0}\in A$ and where $\left(
x_{0},y_{0}\right) $ is a saddle point of \ \bigskip $P\left(
x,y\right) $:}
\[
P\left( x_{0},y\right) \leq \text{\bigskip }P\left(
x_{0},y_{0}\right) \leq \text{\bigskip }P\left( x,y_{0}\right)
\text{ \ \ \ \ }\forall x\in X,\forall y\in Y.
\]

\end{definition}

If this saddle point is not unique, the complex part of $\min \left\{
f\left( z\right) /z\in \Bbb{C}^{n}\right\} $ will be multivalued. It will be
considered that a complex function $f(z)$ is (strictly) \textit{convex} if $%
P\left( x,y\right) $ is (strictly) convex in $x$ and (strictly) concave in $%
y $.

If $f(z)$ is a holomorphic function, then a necessary condition for $z_{0}$
to be a minimum of $f(z)$ on $\Bbb{C}^{n}$is that $f^{\prime }\left(
z_{0}\right) =0$. It is sufficient if $f(z)$ is also convex.

\bigskip
\begin{definition}
\textit{To each complex and convex function, we associate its
\textit{complex Fenchel transform} $\widehat{f}\left( p\right)
:p\in \Bbb{C}^{n}\longmapsto \Bbb{C}$ defined by:}
\[
\widehat{f}\left( p\right) =\max_{z\in \Bbb{C}^{n}}{\left(
p.z-f\left( z\right) \right).}
\]
\end{definition}

Using the classical Lagrange function $L(x,\dot{x},t)$, an
analytical function in $x$ and $\dot{x}$, we define
\cite{Gondran2}, \cite{GondranHoblos} \textit{the complex Lagrange function}
 $L(Z,\mathcal{V},t)$, when replacing $%
x(t)$ by the complex state $Z(t)$\ and $\dot{x}(t)$ by the complex velocity $%
\mathcal{V}\left( t\right) $.

\bigskip
\begin{definition}
\textit{For the process defined by $\left( 1\right) \left(
2\right) $, we define\textit{\ the complex action}
$\mathcal{S}_{\varepsilon }(Z,t)$ using the recurrence equation at
time $t=6q\varepsilon $:}
\[
\mathcal{S}_{\varepsilon }(\widetilde{Z}_{\varepsilon }\left(
t\right) ,t)=\min_{\mathcal{V}(t)}{\frac{1}{6}%
\sum_{j}\left\{ \mathcal{S}_{\varepsilon }(Z_{\varepsilon
}^{j}(t-\varepsilon ),t-\epsilon )+L(\widetilde{Z}_{\varepsilon
}\left( t\right) ,\mathcal{V}(t){\LARGE ,}t)\epsilon \right\}}
\]
\textit{in which the evolution between $Z_{\varepsilon }^{j}(t)$
and $Z_{\varepsilon }^{j}\left( t-\varepsilon \right) $ is given
by the equation (1), and where the min is considered as the
complex minimum for the possible complex velocity
$\mathcal{V}\left( t\right) $. For $t=0$ we take:}
\[
\mathcal{S}_{\varepsilon }\left( Z,0\right) =\mathcal{S}^{0}\left(
Z\right) \text{ \ \ \ \ \ \ }\forall Z\in \Bbb{C}^{3}
\]
\textit{where $\mathcal{S}^{0}\left( Z\right) $ is a given
holomorphic function.}
\end{definition}

\bigskip

This equation (13) can be interpreted as \textit{a new least action
principle adapted to the trajectories of this type}. The decision concerning
the control takes place only at times $t=6q\varepsilon $, i.e. at the times
corresponding to passage into the basic trajectory.

\begin{theorem}
\textit{If the complex process defined by
(1) and (2) has }$L(x,\dot{x},t)=\frac{1}{2}m\dot{x}^{2}-V\left( x\right) $%
\textit{\ as a Lagrangian function, then the complex action
verifies the complex second order Hamilton-Jacobi equation}:
\begin{equation}
\frac{\partial \mathcal{S}}{\partial t}+\frac{1}{2m}\left(
\triangledown \mathcal{S}\right) ^{2}+V\left( Z\right)
-i\frac{\hslash }{2m}\triangle
\mathcal{S}=0\text{ \ \ \ \ \ \ }\forall \left( Z,t\right) \in \Bbb{C}%
^{3}\times \Bbb{R}^{+}
\end{equation}

\begin{equation}
\mathcal{S}\left( Z,0\right) =\mathcal{S}^{0}\left( Z\right)
\text{ \ \ \ \ \ \ }\forall Z\in \Bbb{C}^{3}
\end{equation}
\end{theorem}

\textit{Proof: } For the proof, we suppose that
$\mathcal{S}_{\varepsilon }(Z,t)$ is a very smooth function on
$\varepsilon$ and holomorphic at $Z$ and C$^{1}$ at $t.$ By the
lemma 2.10, we have

\[
\frac{1}{6}\sum_{j}\left\{ \mathcal{S}_{\varepsilon
}(Z_{\varepsilon }^{j}(t-\varepsilon )\right\}
=\mathcal{S}_{\epsilon }(\widetilde{Z}\left( t\right)
,t)-D\mathcal{S}_{\varepsilon }(\widetilde{Z}\left( t\right)
,t)+o(\varepsilon) .
\]

Whence, we deduce at point $\left( Z,t\right) $ the following equation:

\begin{equation}
\frac{\partial \mathcal{S}_{\epsilon }}{\partial
t}=\min_{\mathcal{V}}{\left( L(Z,\mathcal{V}{\LARGE
,}t)-\mathcal{V}\cdot
\triangledown \mathcal{S}_{\epsilon }+i\frac{\hslash }{2m}\triangle \mathcal{%
S}_{\epsilon }+o\left( \epsilon \right) \right)}.
\end{equation}
\ Using complex Fenchel transform of $L(Z,\mathcal{V},t)$ in
$\left( 16\right) $ and doing $\epsilon \rightarrow $ 0$^{+},$ we
obtain $\left( 13\right) $.$\Box $

\bigskip

\bigskip If we take for wave function $\Psi =e^{i\frac{\mathcal{S}}{\hslash }%
}$ and apply the restriction of $\left( 14\right) \left( 15\right)
$ to the real part of Z , the theorem 3.4 becomes:

\bigskip
\begin{theorem}
\textit{If the complex process defined by
(1)(2) has }$L(x,\dot{x},t)=\frac{1}{2}m\dot{x}^{2}-V\left( x\right) $%
\textit{\ for Lagrangian function, then the wave function }$\Psi $ \textit{%
verifies the Schrödinger equation}:
\[
i\hslash \frac{\partial \Psi }{\partial t}=\mathcal{-}\frac{\hslash ^{2}}{2m}%
\triangle \Psi +V(X)\Psi \qquad \forall (X,t)\in \Bbb{R}^{3}\times \Bbb{R}%
^{+}
\]

\[
\Psi (X,0)=\Psi ^{0}(X)\qquad \forall X\in \Bbb{R}^{3}.\Box
\]
\end{theorem}

\bigskip

As $L(Z,\mathcal{V},t)=\frac{1}{2}m\mathcal{V}^{2}-V\left( Z\right) $, the
minimum of (15) is obtained with $m\mathcal{V-\nabla S}%
_{\varepsilon }=0$. Then we have
\begin{equation}
\mathcal{V}\left( t\right) =\frac{\triangledown \mathcal{S}}{m}
\end{equation}

By breaking down $\mathcal{S}(X,t)$ into its real and imaginary parts, $%
\mathcal{S}(X,t)=S(X,t)-\frac{i\hslash }{2}\ln \rho (X,t)$, because the wave
function is also written as $\Psi =\rho e^{i\frac{S}{\hslash }}$, we can
deduce that the real basic trajectory $\widetilde{X}(t)$ verifies the
classical differential equation:
\begin{equation}
\frac{d\widetilde{{\large X}}(t)}{dt}=\frac{\nabla S}{m},\text{ \ \ \ \ \ \
\ \ \ }\widetilde{{\large X}}(0)={\large X}_{0}.
\end{equation}

This is the trajectory proposed by de Broglie $\left[ 4\right] $ and Bohm $%
\left[ 1\right] $.

\bigskip
\begin{theorem}
\textit{If the complex process defined by
(1) and (2) has }$L(x,\dot{x},t)=\frac{1}{2}m\dot{x}^{2}-V\left( x\right) $%
\textit{\ for Lagrangian function, then the} \textit{real part of
the basic trajectory follows the trajectory proposed by de Broglie
and Bohm.}

\end{theorem}

\bigskip

A fundamental property of this trajectory is that the density of probability
$\varrho (x,t)$ of a family of particles satisfying (18), and having a
probability density $\rho _{0}\left( x\right) $ at initial time, verifies
the Madelung continuity equation:
\begin{equation}
\frac{\partial \varrho }{\partial t}+div(\varrho \frac{\nabla S}{m})=0
\end{equation}
so that the trajectories are consistent with the Copenhagen interpretation,
cf. by exemple $\left[ 5\right] $.

\bigskip
\begin{remark}
The most natural hypothesis for the choice of $%
\varepsilon $ is to link it to the de Broglie wavelength. Now, the
internal motion of the process defined by $\left( 1\right) $ and
$\left( 2\right) $ has a period of $6\varepsilon .$ Thus, it is
possible to identify this period to the frequency of de Broglie
and to put:

\begin{equation}
6\varepsilon =T=\frac{\lambda _{dB}}{v}=\frac{h}{mv^{2}}
\end{equation}
\end{remark}

\section{Conclusion}

There are some questions about this model. What is the sense of
the imaginary velocity? Is it the good model for the Schrödinger
equation?

\bigskip

The complex velocity $\mathcal{V}\left( t\right) $ of the process,
given by (16), is written as:
\begin{equation}
m\mathcal{V}\left( t\right) =\nabla S-i\nabla \log \rho \frac{\hbar }{2}.
\end{equation}
The original velocity proposed by de Broglie and Bohm is the real
part $v(t)$ of \ $\mathcal{V}\left( t\right) $%
\begin{equation}
mv\left( t\right) =\nabla S.
\end{equation}

However, it is possible to show, cf.\cite{Holland2} and
\cite{GondranGondran}, that for particles with a constant spin
\textbf{s}, as it is the case in the Schrödinger approximation,
the Dirac equation implies that the momentum of a particle must be
given by:
\begin{equation}
m\mathcal{V}\left( t\right) =\nabla S+\nabla \log \rho \times
\mathbf{s}
\end{equation}
where the spin-dependant current is the Gordon current. The
equation (21) is relevant only to spin-0 particles. This
spin-dependant term was often suggested, cf. \cite{Bohm2},
\cite{Holland1}, but only in the context of the Pauli equation and
not the Schrödinger equation. This term is naturally obtained from
Dirac equation and this representation in a Clifford algebra or
the quaternion algebra. The momentum given by (22) has been
recently valided, cf. \cite{Colijn} and \cite{GondranGondran}, for
hydrogen eigenstates.

Because $\mathbf{s=}\frac{\hbar }{2}\mathbf{u}$ (with $\mathbf{u}$
unitary), we can consider the equation (20) as the projection on
the complex field of the equation (22) where
\[
\nabla S+\nabla \log \rho \times \mathbf{s}
\]
is the gradient of the quaternion
\[
S+log\rho \mathbf{s}
\]
as
\[
\nabla S-i\nabla \log \rho \frac{\hbar }{2}
\]
is the gradient of the complex number
\[
S-i\log \rho \frac{\hbar }{2}.
\]

We can conclude that the process defined by $\left( 1\right) $ and
$\left( 2\right) $ is only a first approximation on the complex
field of a more general model, certainly based on a Clifford
algebra or the quaternion algebra.

\bigskip





\bigskip

\begin{thebibliography}{99}
\bibitem{Bohm}
Bohm D.J., 1952: A suggested interpretation of the quantum theory
in terms of"hidden" variables. \textit{Physical Review},
\textbf{85,} 166-193.

\bibitem{Bohm2}
Bohm D., and Hiley B.J., 1993: \textit{The Undivided Universe}.
Routledge, London and New York.

\bibitem{Colijn}
Colijn C., and Vrscay E.R., 2002: Spin-dependent Bohm trajectories
for hydrogen eigenstates. \textit{Physics Letters A} \textbf{300,}
334-340.

\bibitem{Broglie}
de Broglie, 1927: La mécanique ondulatoire et la structure de la
matière et du rayonnement. \textit{Le Journal de Physique et le
radium}. série 6, Vol. 8, n$^\circ 5$, 225-241.

\bibitem{Espagnat}
d'Espagnat B., 1983: \textit{In Search of Reality}. Springer,
New-York.

\bibitem{Feynman}
Feynman R.P., and Hibbs A.R., 1965: \textit{Quantum Mechanics and
Path Integrals}. Mc Graw-Hill, New York.

\bibitem{Gondran2}
Gondran M., 2001: Calcul des variations complexe et solutions
explicites d'équations d'Hamilton-Jacobi complexe.
\textit{C.R.Acad.Sci.}, Paris, \textbf{332}, sérieI, 677-680.

\bibitem{Gondran3}
Gondran M., 2001: Processus stochastique non standard en
mécanique. \textit{C.R.Acad.Sci.}, Paris, \textbf{333}, sérieI,
593-598.

\bibitem{GondranHoblos}
Gondran M., and Hoblos R., 2003: Complex calculus of variations.
\textit{Kybernetika Max-Plus special issue}, \textbf{39,} number
2.

\bibitem{GondranGondran}
Gondran M., and Gondran A., 2003: Revisiting the Schrödinger
Probability current. quant-ph/0304055v1.

\bibitem{Holland1}
Holland P.R., 1993: \textit{The quantum Theory of Motion.}
Cambridge University Press.

\bibitem{Holland2}
Holland P., 1999: Uniqueness of paths in quantum mechanics. Phys.
Rev.A \textbf{60} 4326.

\bibitem{Maslov}
Maslov V.P., 1987: \textit{Méthodes opérationnelles}. Edition Mir,
Moscou.

\bibitem{Nelson1}
Nelson E., 1966: Derivation of the Schrödinger Equation from
Newtonian Mechanics. \textit{Physical Review}, \textbf{150}, n°4,
1079-1085.

\bibitem{Nelson2}
Nelson E., 1985: \textit{Quantum Fluctuations}. Princeton
University Press, Princeton, N.J..

\bibitem{Nottale}
Nottale L., 1993: \textit{Fractal Space-Time and Microphysics,
Towards a Theory of Scale Relativity}. World Scientific
(Singapore, New Jersey, London).

\bibitem{Robinson}
Robinson A., 1973: Function theory on some non-archimedean fields.
\textit{Amer. Math. Monthly}, \textbf{80,} 87-109.

\end{thebibliography}
\end{document}